# On Multi-domain Network Slicing Orchestration Architecture & Federated Resource Control

Tarik Taleb†, Ibrahim Afolabi†, Konstantinos Samdanis‡, and Faqir Zarrar Yousaf±
† Aalto University, Espoo, Finland
‡ Nokia-Bell Labs, Munich, Germany
± NEC Laboratories Europe GmbH, Heidelberg, Germany

*Abstract*—A sophisticated and efficient network slicing architecture is needed to support the orchestration of network slices across multiple administrative domains. Such multi-domain architecture shall be agnostic of the underlying virtualization and network infrastructure technologies. Its objective is to extend the traditional orchestration, management and control capabilities by means of models and constructs in order to form a well-stitched composition of network slices. To facilitate such composition of networking and compute/storage resources, this paper introduces a management and orchestration architecture that incorporates Software Defined Networking (SDN) and Network Function Virtualization (NFV) components to the basic 3GPP network slice management. The proposed architecture is broadly divided into four major strata, namely Multi-domain Service Conductor Stratum, Domain-specific Fully-Fledged Orchestration Stratum, Sub-Domain Management and Orchestration (MANO) and Connectivity Stratum, and Logical Multidomain Slice Instance stratum. Each of these strata is described in detail providing also the fundamental operational specifics for instantiating and managing the resulting federated network slices.

## I. Introduction

The 5$^{th}$ Generation of Mobile Networks (5G) is envisioned to revolutionize the communication service experience, enabling also new applications. It is expected to offer content-rich multimedia in a crowd and on the move, support critical communications and allow massive connectivity of sensors and actuators [1]. Such a plethora of services would accelerate emerging business opportunities. It facilitates commercialization for vertical segments without a network infrastructure by utilizing customized networks and cloud resources. Indeed, 5G introduces the concept of network slicing, which is based on virtualization and softwarization. Network slicing enables programmability and modularity in the provisioning of network resources with respect to specific vertical segment service requirements, thereby advancing the apriori 4G monolithic architecture [2]. Typically, different verticals offer applications with distinct and often conflicting service requirements in terms of bandwidth, latency, etc. Allowing a variety of verticals to use a common infrastructure, requires an appropriate level of isolation and QoS provisioning. This can only be addressed via an efficient means of resource orchestration and programmable management [3].

Network slices, allocated to verticals, can stretch across greater geographical areas, i.e. between different countries, or encompass areas where coverage can only be assured by combining resources from different mobile operators. Likewise, vertical services may need computing and storage resources that can only be offered by particular cloud providers to complement networking capabilities. Such slice deployment requires an efficient combination of federated resources. Resources not only to provide the desired bandwidth, but to also cope with additive constraints (e.g., latency or jitter) and multiplicative constraints (e.g., end-to-end error rate probability) across multiple administrative domains.

Tarik Taleb is also with the Faculty of Information Technology and Electrical Engineering, Oulu University, and with the Department of Computer and Information Security, Sejong University, Seoul 05006, South Korea.

Fulfilling such requirements across a federated environment is challenging, not only from the perspectives of decomposing a slice request into respective domain(s), but also assuring its performance maintenance. This paper proposes a multi-domain network slicing orchestration architecture introducing the notion of Multi-domain Service Conductor stratum, which provides service management across federated domains. The Multi-domain Service Conductor stratum analyzes and maps the service requirements of incoming multi-domain slice requests onto the respective administrative domains. It also maintains the desired service performance throughout the entire service life-cycle. To handle the dynamics related to federated resource allocation efficiently, a crossdomain coordinator is introduced. Such cross-domain coordinator aligns cloud and networking resources across federated domains and carries out the life-cycle management (LCM) operations of a multi-domain slice. It also establishes and controls inter-domain transport layer connectivity assuring the desired performance.

The remaining of this paper is organized as follows. Section II presents the fundamentals of network slicing, highlighting the challenges for slice management in a multi-domain federated environment. In view of these challenges and gaps, the details of our proposed multidomain slice management framework are presented in Section III. Section IV describes the main procedures, while a discussion on open challenges is presented in Section V. Finally Section VI concludes the paper.

## II. Network Slicing Concepts, Key Enablers and Open Issues

### *A. Network Slicing in Single Administrative Domains*

A network slice (NS) is a fundamental but complex attribute of a 5G network. According to NGMN [4], an NS is defined as *a set of network functions, and resources to run these network functions, forming a complete instantiated logical network to meet certain network characteristics required by the Service Instance(s)*. In other words, an NS is a basic substrate offered by mobile network operators from which multiple business services are deployed and run in an efficient and costeffective manner. Virtualization is the key technology enabler for realizing a fully or partially isolated NS Instance (NSI). It abstracts the physical and/or virtual infrastructure resources, such as computation, network, memory, and storage, offering logical resources with customized policy and configuration parameters. Such logical resources are then assigned to different tenants, e.g., verticals, which form virtualized functions and overlay connectivity fulfilling the desired service needs.

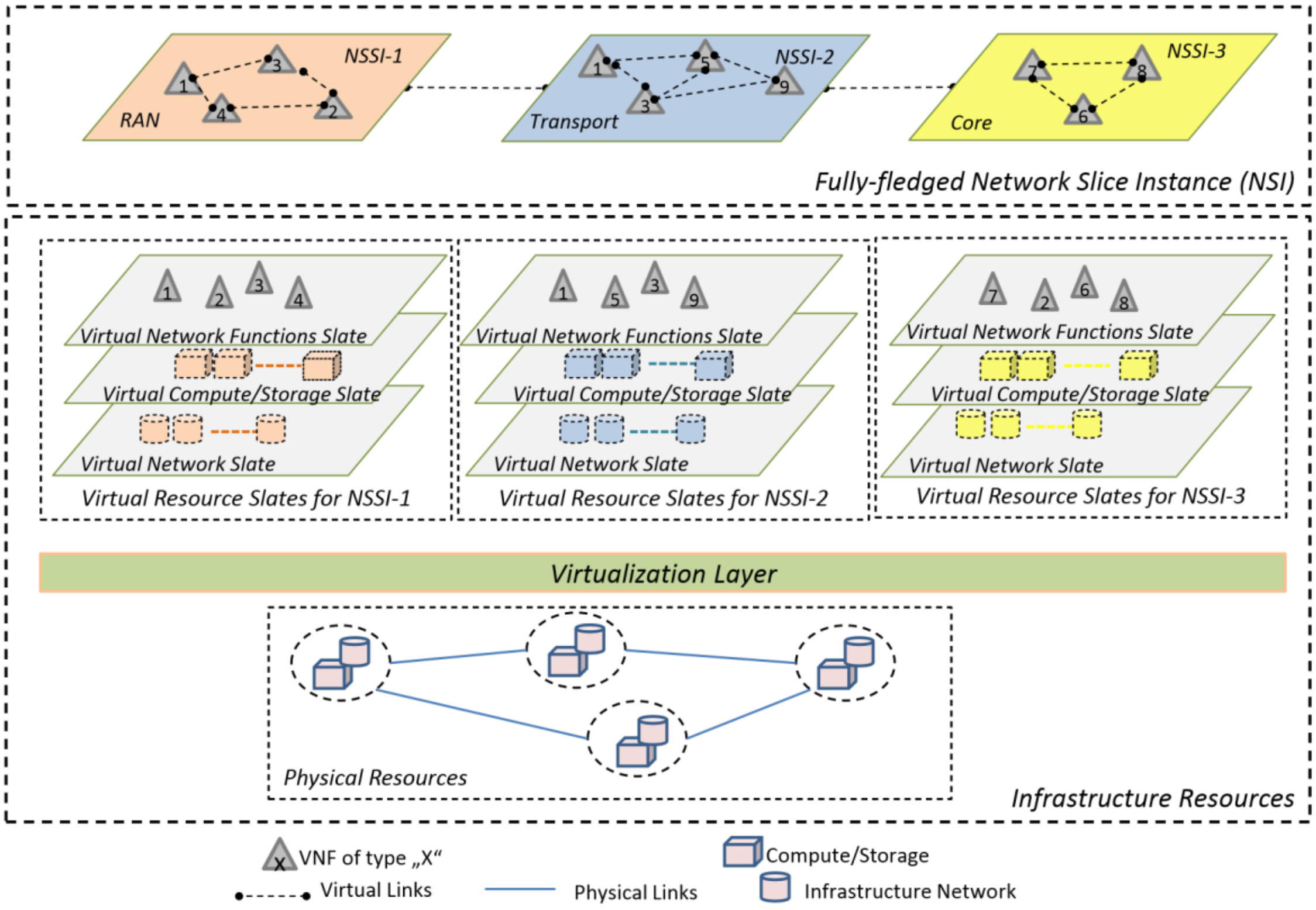

Figure 1: Composition of a Fully-fledged Network Slice Instance

A Virtualized Network Function (VNF) can accommodate simple network functions, e.g., virtual firewall, or more complex ones, e.g., virtual mobile core network [1]. Each VNF is assigned a specific type and amount of virtualized-resources. VNFs are interconnected in a overlay network topology order over well-defined virtual or logical links to create a *Fully-Fledged* NSI. A NSI typically consists of multiple Network Slice Subnet Instances (NSSIs) that represent a group of network function instances and/or logical connectivity. An NSSI forms a part or complete constituents of an NSI [5].

Figure 1 illustrates the concept and composition of a Fully-Fledged NSI, which consists of three NSSIs, each belonging to a different technology domain, e.g., Radio Access Network (RAN), transport and core. The RAN and core NSSIs are composed of VNF(s) interconnected over logical transport links. A NSSI can, on its own, be a Fully-Fledged NSI, but then multiple NSSIs stitched together can extend the service scope and thus create a new enhanced Fully-Fledged NSI. Each NSSI is established over an infrastructure that provides virtualized resources, which are accompanied by the appropriate orchestration and management means forming the socalled *slate*, as shown in Figure 1. Each resource slate abstracts a particular type of resource and provides the means of LCM and control towards the NSSI.

For example, a virtual compute or storage slate is characterized by virtual compute resources such as virtual Central Processing Unit (vCPU), virtual memory, and virtual storage capacity. Resource slates at the network edge can reduce latency by offering users combine caching and computation offloading capabilities [6]. Equivalently, a virtual connectivity slate is characterized by virtualized network resources, such as virtual network interface controller, logical links, virtual switches etc. Each NSSI is assigned a set of resource slates, the type and amount of which depends on the requirements of the VNFs that are part of the NSSI. Resource slates shall include an interface and data model, e.g., based on YANG, that offer the means for providing LCM to NSSI directly or through the corresponding orchestrator and/or controller. Currently, proprietary interfaces may be used for this purpose until standardized solution are developed.

A comprehensive and well-coordinated management system is required to facilitate an effective LCM of a Fully-Fledge NSI. At minimum, the following management, orchestration and control entities are essential:

- *Network Slice Manager* responsible for the configuration and operation of a mobile network service to a Fully-Fledge NSI.
- *NFV MANO*[2] that instantiates and orchestrates the requested VNFs considering the supported availability.
- *SDN Controller* that connects together VNFs forming service function chains and controls the transport layer connectivity.

3GPP has introduced an orchestration and management architecture in [5] consisting of: (i) a service management function that analyzes incoming slice requests, converting service requirements into networking ones and (ii) a network slice management function, which performs the mapping onto network resources and takes care of the LCM. Although the resource mapping process is carried out across different technology domains, including the RAN, transport and core, the current 3GPP efforts concentrate only on NSIs deployed and managed by a single administrative entity.

### B. Multi-domain Slice Management

An end-to-end NS is deployed across multiple networks, stretching across the RAN, transport and core network segments; belonging to the same or different administrative domains. The process of establishing a multi-domain NSI leverages the benefits of recursive virtualization as described in [7]. Recursive virtualization allows a hierarchical network abstraction, wherein slates offer a logical resource view to NSSIs, and NSSIs in turn to the NSI. Each successively highest level enables a greater abstraction within a broader scope hiding the layer internals, while allowing a generic resource usage. Such a paradigm can easy the composition of NS across different administrative borders, combining efficiently and in a flexible manner different types of resources.

The main challenge is laid on the *deployment* and *runtime management* since the involved domains may not only be geographically apart data centers interconnected over a Wide Area Network (WAN) infrastructure, but may belong to different administrative domains. Figure 2 depicts a fully functional end-to-end

[1] like the virtual Evolved Packet Core (vEPC)

[2] ETSI, Network Functions Virtualisation (NFV); Management and Orchestration, GS NFV-MAN 001 v1.1.1, Dec. 2014.

NS across three administrative domains A, B and C, illustrating the respective physical infrastructure. A multi-domain NSI shall combine two or more Fully-Fledged NSIs that belong to different administrative domains facilitating an end-to-end multi-domain (*a.k.a* federated) NSI. The constituent Fully-Fledged NSIs instantiated from the different administrative domains are also referred to as NSSI of the multi-domain NSI.

To understand the inherent complexity, we list below some of the main processes involved when a slice request is received from a 3$^{rd}$ party:

1) Mapping of the service requirements onto capability requirements.
2) Translating the capability requirements into:
    a) NSI resource requirements in terms of compute, storage and networking resources.
    b) NSI topology and connectivity type, policy, isolation and security requirements.
3) Identifying the infrastructure-domains with the required resources, which can assure the end-to-end NSI functional and operational requirements.
4) Instantiating NSSIs in each infrastructure domain and then "stitching" them to create the federated NSI.
5) Providing run-time coordination management operations across different domains for maintaining the end-to-end NSI service integrity.

Such processes have several architectural implications requiring effective controllers at different stratum levels. As shown in Figure 2, there is a need to have at least three levels of controllers. That is, two sub-domain controllers for the orchestration of Network Function Virtualization Infrastructure (NFVI) resources and networking control. A domain-specific slice controller for the orchestration and management of NSSIs within each respective domain, and an overarching end-to-end slice coordinator for unifying the management of individual multi-domain NSI.

A preliminary study towards a framework for virtualization across multiple administrative domains is introduced in [8] elaborating the main concepts of isolation, programmability and performance

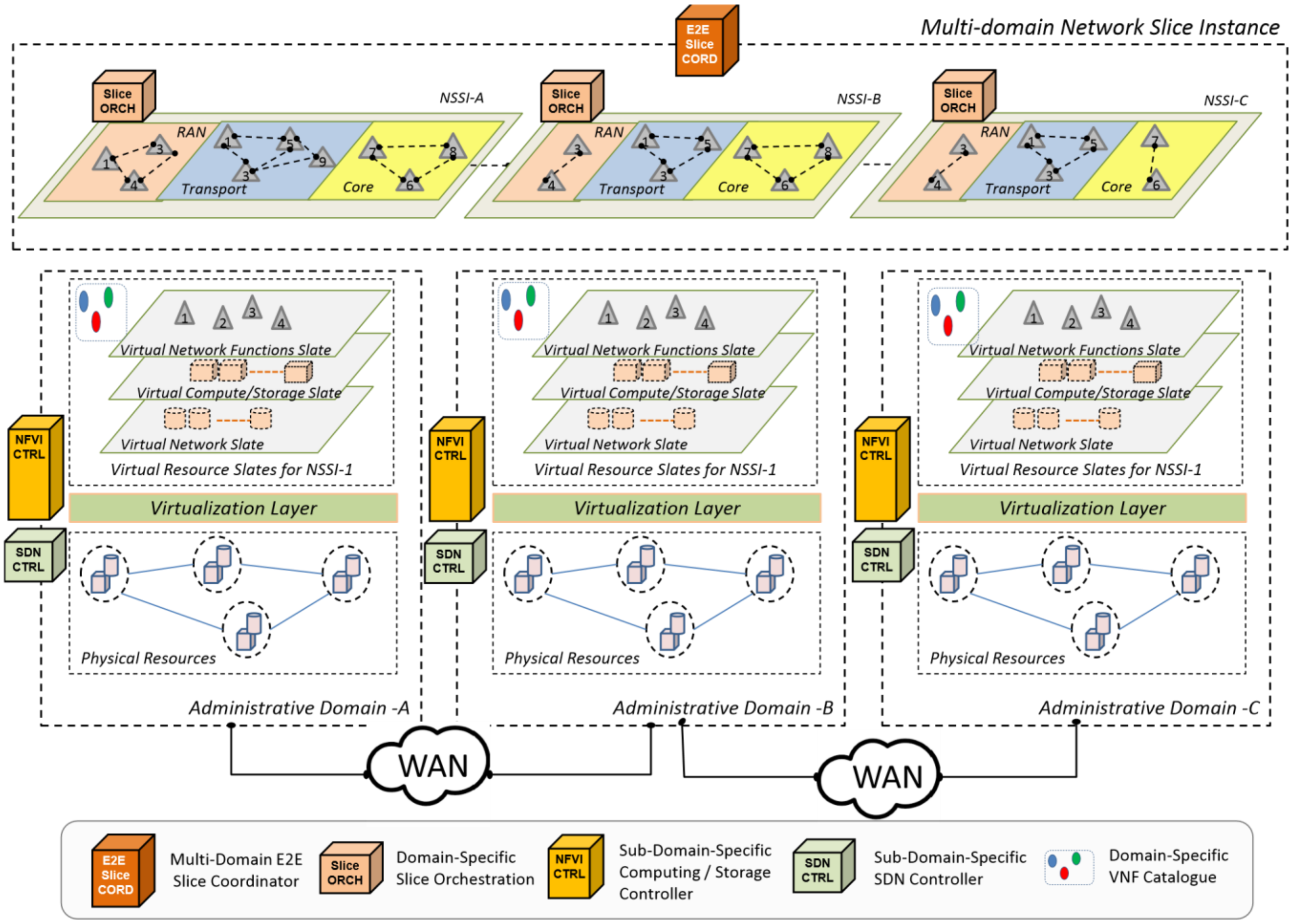

Figure 2: Federated Network Slice Across Multiple Administrative Domains

maintenance, also including the fundamental functional components.

Logical resources from different administrative domains are collected by a virtualization resource manager, which acts as a broker allowing third parties to establish a virtual network optimized for supporting particular services. A federated slicing solution is presented in [9] introducing the notion of multi-domain orchestrator, which handles slice requests for resources beyond its domain. The proposed multi-domain orchestrator analyzes the related service requirements and directly contacts the appropriate neighboring domains performing resource negotiation. Once a slice is established, a peer-to-peer management plane is responsible for handling the LCM considering relevant service-oriented key performance indicators, while coordinating closely with individual domain-specific orchestrators.

A hierarchical multi-domain orchestration architecture is introduced in [10], based on the concept of recursive abstraction and resource aggregation that "stitches" NSI heterogeneous resources initially on per domain level and then across federated domains. A similar concept is presented in [11] where an overarching Inter-slice Resource Broker functional element is proposed to manage and orchestrate resources for end-to-end slices across multiple technology domains. Each domain facilitates a local instance of the standard ETSI NFV-MANO interacting with the broker. Although different technology domains may belong to a distinct administration, the solution assumes a unified orchestration and management provided by a single administrative domain. Such unified orchestration and management acts as aggregator without supporting service federation to form an end-toend multi-domain NSI.

With regard to multi-domain support, ETSI NFV to date has published two informative reports. The first report [12] deals with managing the connectivity of an NS deployed over multiple NFVI sites, referred to as NFVI-PoPs. A single MANO system then manages interconnectivity issues over WAN links linking these NFVIPoPs. The second report [13] highlights the different architectural options and recommendations to support MANO operations in multiple administrative domains. To manage multi-site/multi-domain NSIs, a direct reference point between the NFV Orchestrator (NFVO) functional elements is recommended in each NFVI-PoP. Such a peer-to-peer approach does not only bring more complexity, but is not optimal in view of the delay sensitive nature of MANO operations. In view of the challenges and gaps discussed above, we present a novel multi-domain architecture for the LCM of NSIs deployed across heterogeneous federated infrastructure domains.

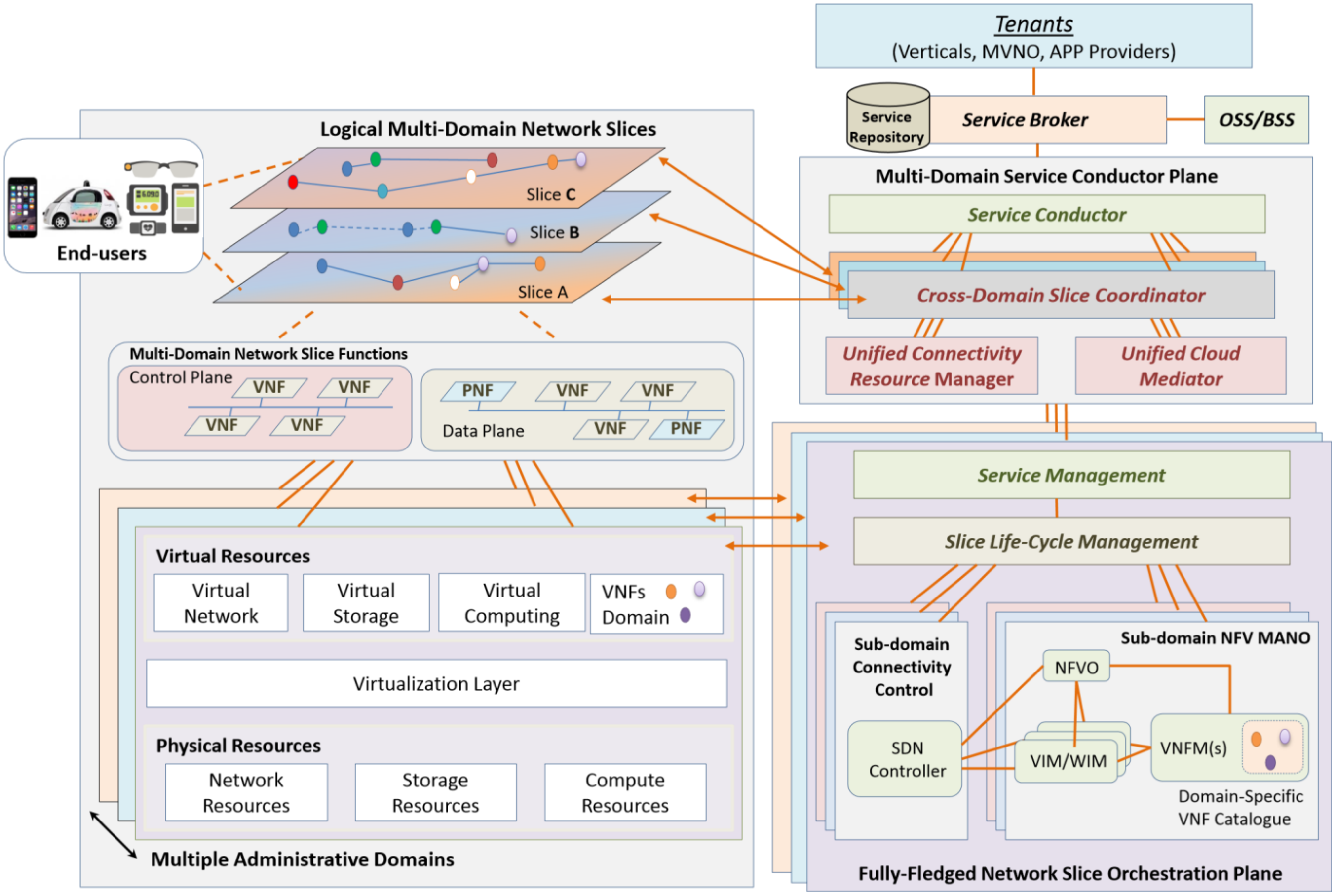

Figure 3: Multi-domain Slicing Architecture.

## III. Multi-domain Orchestration Architecture

The envisioned architecture for NS across multiple administrative and technological domains is illustrated in Figure 3. The architecture takes into account scalability, with its fundamental components and functions described below.

### *A. Service Broker Stratum*

The envisioned architecture introduces a functional plane known as Service Broker [14] to handle incoming slice requests from verticals, Mobile Virtual Network Operators (MVNO), and application providers, with the main operations listed below:

- NS admission control and negotiation considering service aspects.
- Management of slice user/owner relationship enabling a direct tenant interface with the federated Multi-domain Service Conductor plane.
- NSI revenue management, which involves billing and charging of slice owners.
- NSI scheduling, i.e. start and termination time related with slice composition and decommission.

Typically, a Service Broker collects abstracted service capability information regarding different administrative domains, creating a global service support repository. It also interacts with the Operating/Business Support System (OSS/BSS) in order to collect business, policy and administrative information when handling slice requests.

### *B. Multi-domain Service Conductor Stratum*

The Multi-domain Service Conductor Stratum is responsible for service orchestration and management across federated resources related with successfully admitted slice requests. It consists of the following two main building blocks:

- *Service Conductor* that decomposes a slice request towards different administrative domains and decides on the combination of domains, including also the cross-domain connectivity, i.e. “stitching”. It instantiates the Cross-domain Slice Coordinator with respect to a particular federated NSI to perform LCM processes and assigns ownership rights, e.g., offered to a vertical or MVNO. A Service Conductor also carries out potential service specific re-adjustments across the federated domains, i.e., instantiating, modifying and decommissioning domains, upon request in case of performance degradation or service policy update.
- *Cross-domain Slice Coordinator* monitors, manages and controls the corresponding resources related with a federated NSI, while ensuring secure and trusted connectivity across administrative domains. It also serves as a mediator among federated resources, carrying out domain specific resource allocation and re-adjustments to compensate potential performance degradation. A Cross-domain Slice Coordinator performs federated compute, storage and network resource allocation with the help of:
    1) Unified Cloud Mediator that interprets and translates the performance capability description of heterogeneous cloud resources.
    2) Unified Connectivity Resource Manager, which negotiates cross-domain connectivity across different administrative domains.

When a federated NSI is formed, the Multi-domain Service Conductor Stratum achieves scalability by assigning a Cross-domain Slice Coordinator that can independently point out domain and cross-domain service misbehaviors accurately for each NSI.

### *C. Fully-Fledged Network Slice Orchestration Stratum*

The Fully-Fledged Network Slice Orchestration Stratum interacts with the Cross-domain Slice Coordinator, allocating internal domain resources for establishing a federated NSI. It also provides the corresponding LCM via the following functional blocks:

- *Service Management Function* analyzes the slice request received from the Cross-domain Slice Coordinator and identifies the RAN and core network functions, including value added services. It also determines logical links characterized by bandwidth, delay, jitter, packet loss, etc. In return, it feeds the Cross-domain Slice Coordinator with service and performance capability information related with the underlying resources.
- *Slice Life-Cycle Management Function* identifies the appropriate network slice template from an associated catalogue and forms a logical network graph. Such graph is mapped to the underlying compute, storage and network resources corresponding to a technology specific slate. For deploying particular slates, further information (e.g., the desired topology type such as multi-cast tree, policy and control plane functions) can be provided. The Slice LifeCycle Management Function is also responsible for the instantiation, run-time and orchestration of a NSSI considering the resource slates within the same administrative domain, performing monitoring and modification related operations.

- *Sub-domain NFV MANO* takes care of the VNF, computation or storage slates. It communicates with the Slice Life-Cycle Management Function providing an abstracted view of the underlying infrastructure and performs the instantiation and run-time operations of the corresponding VNF, computation or storage slates with the assistance of the following functional blocks:
    1) NFVO is aware of the LCM related with the share of virtual resources apportioned to each slate under its control. It is the decision making entity for the allocation of available virtual resources, which is periodically reported by the corresponding VIM/WIM and VNF Managers (VNFMs).
    2) VNFM is in charge of the LCM (i.e., instantiating, monitoring, modifying, and terminating) VNFs. In collaboration with the NFVO, it is responsible for allocating the optimal amount of resources to particular VNFs and for handling dynamic VNF re-configurations based on received updates.
    3) VIM/WIM is carrying out resource management functionalities, interacting also with the VNFs and virtualized network infrastructure. To simplify its deployment and enhance its modularity, VIM/WIM consists of:
        a) Virtual Network Resources Manager that directly controls the networking resources within the virtualized environment.
        b) Virtual Compute Resources Manager works closely with the NFVI storage and computing controller scaling up or down of virtual machine’s CPU resources.
        c) Virtual Storage Resources Manager is responsible for allocating the appropriate amount of virtual storage resources, abstracted directly from the NFVI.
- *Sub-domain SDN Controller* provides the network connectivity and service chaining among the allocated VNFs connecting remote cloud environments optionally via Physical Network Functions (PNFs), e.g., routers or switches. It feeds the Slice LifeCycle Management Function with an abstracted network resource view and monitoring reports for assuring the desired Service Level Agreement (SLA) in case of a failure or performance degradation. The Sub-domain SDN Controller can leverage the benefits of deep data plane programmability and information centric networking for the transport layer. For integrating the Sub-domain SDN Controller with the NFV MANO architecture, two options were considered as documented in [15]. The SDN Controller being: (i) a part of VNFI interacting with the VIM/WIM or (ii) an independent PNF entity linked with the NFVO, (typically via the corresponding sub-domain OSS/BSS). The former suits better a virtualized environment, while the later a mixed, including PNFs.

### *D. Sub-domain Infrastructure Stratum*

The Sub-domain Infrastructure Stratum consists of the physical and virtual infrastructure containing:

- VNFs that are related with the hardware infrastructure from where they can be deployed. These network functions are typical 5G control plane and data plane functions, or value added services such as a firewall or Content Delivery Network (CDN).
- Virtual resources are the abstracted physical resources that VNFs are running directly on, i.e., the virtual compute, storage and networking resources.
- Virtualization layer often referred to as the hypervisor that sits directly above the physical infrastructure is responsible for partitioning the physical resources among the operating VNFs. It also abstracts the underlying hardware resources and decouples VNFs from hardware.
- Physical Infrastructure consists of the hardware resources that provide processing, storage and network connectivity functionalities to the VNFs through the virtualization layer. The computing and storage physical resources are usually Commercial

Table I: Network Slicing Orchestration Architectures and their Offered Support.

| Orchestration Architectures | Multi Admin. Domains | Multi Tech. Domains | RAN Orch. | Broker: AC/Neg. | Service Chain & SDN | Service Mang. | Federated LCM | $3^{rd}$ Party Cntl./Orch. | Program-ability | Recursive Virtualiz. | Unif. Connct. Mgmt. | Unif. Cloud Med. |
|---|---|---|---|---|---|---|---|---|---|---|---|---|
| 3GPP 28.530 [5] | 7 | 3 | 3 | 7 | 7 | 3 | 3 | 7 | 7 | 7 | 7 | 7 |
| SDN TR-526 [7] | 3 | 7 | 7 | 7 | 3 | 3 | 3 | 3 | 3 | 3 | 7 | 7 |
| ITU-T Y.3011 [8] | 3 | 3 | 7 | 3 | 7 | 3 | 3 | 3 | 3 | 3 | 3 | 7 |
| 5G-EX [9] | 3 | 3 | 7 | 7 | 7 | 3 | 3 | 7 | 3 | 7 | 7 | 3 |
| 5G!Pagoda [10] | 3 | 3 | 3 | 7 | 3 | 7 | 3 | 7 | 3 | 7 | 3 | 3 |
| 5G-Norma [11] | 7 | 3 | 3 | 3 | 3 | 7 | 7 | 3 | 3 | 7 | 7 | 7 |
| NFV-MANO [13] | 3 | 3 | 7 | 7 | 7 | 7 | 3 | 7 | 3 | 7 | 7 | 3 |
| Proposed MDO | 3 | 3 | 3 | 3 | 3 | 3 | 3 | 3 | 3 | 3 | 3 | 3 |

Off The Shelf (COTS) commodity servers with general purpose CPUs and local or network attached hard disk storage. Network resources are typically physical switches and routers, but virtualized counterparts are also considered.

### *E. Multi-domain Orchestration: A Quantitative Analysis*

To highlight and distinct the features and operations of our architecture a quantitative analysis and comparison is provided. Other approaches considered include standard efforts based on 3GPP, ONF-SDN, ITU-T and ETSI NFV-MANO and representative research projects focusing on 5G-EX, 5G!Pagoda and 5G-NORMA. To make the analysis concise, the following set of features are selected:

- Multi-domain support: Multiple administrative domains and technology types including RAN
- Multi-domain service and resource management: Service broker, service management and federated LCM
- Multi-domain tenant control: $3^{rd}$ party NS control/orchestration, programmability and recursive virtualization
- Multi-domain resource "'stiching": Unified multidomain connectivity and cloud mediation

Table I summarizes the details of the quantitative analysis, showing the functional and operational features where our proposed multi-domain architecture advances the state of the art.

## IV. Multi-domain Network Slice Orchestration & Management Procedures

To further expatiate on the orchestration of multidomain network slicing, a series of operational procedures are elaborated considering slice configuration and modification.

### *A. Multi-domain Network Slice Configuration*

A multi-domain NSI slice is instantiated following the procedure illustrated in Figure 4. A slice request first arrives at the Service Broker, which performs the admission control and negotiation with the requesting tenant considering the OSS/BSS policy and rules. Successful requests are forwarded to the Service Conductor, which analyzes the service requirements selecting the appropriate domains before instantiating a Cross-domain Slice Coordinator for the newly allocated multi-domain NSI. The Service Conductor or optionally the requesting tenant once authorized, programs the Cross-domain Slice Coordinator providing essential information related to the desired service type (e.g., SLA and policy).

### *B. Multi-domain Network Slice Modification*

Once the Cross-domain Slice Coordinator is configured, the Service Conductor provides the corresponding service decomposition details of the slice request. The Cross-domain Slice Coordinator relies on the Unified Cloud Mediator for guidance on interpreting the slice requirements related with VNFs and value added services across heterogeneous platforms. Cross-domain connectivity is established through the Unified Connectivity Resource Manager. Thereafter, the Cross-domain Slice Coordinator establishes a secure communication with each Service Management Function in the relative administrative domain. It then provides service type specifics (e.g., SLA and policy) related to the the relevant VNFM. When the request directly reaches the VIM, it represents a situation of resource scaling related with a shared VNF resource. However, requests for instantiating VNFs are handled by the Subdomain VNFM. For the connectivity slate, the Subdomain SDN Controller performs the necessary network configurations to establish the transport layer and corresponding slice request. Each Service Management Function in turn performs a mapping analysis to identify the network resources, i.e. network functions, value added service and connectivity, that correspond to certain technology sub-domains and then informs the Slice LifeCycle Management Function.

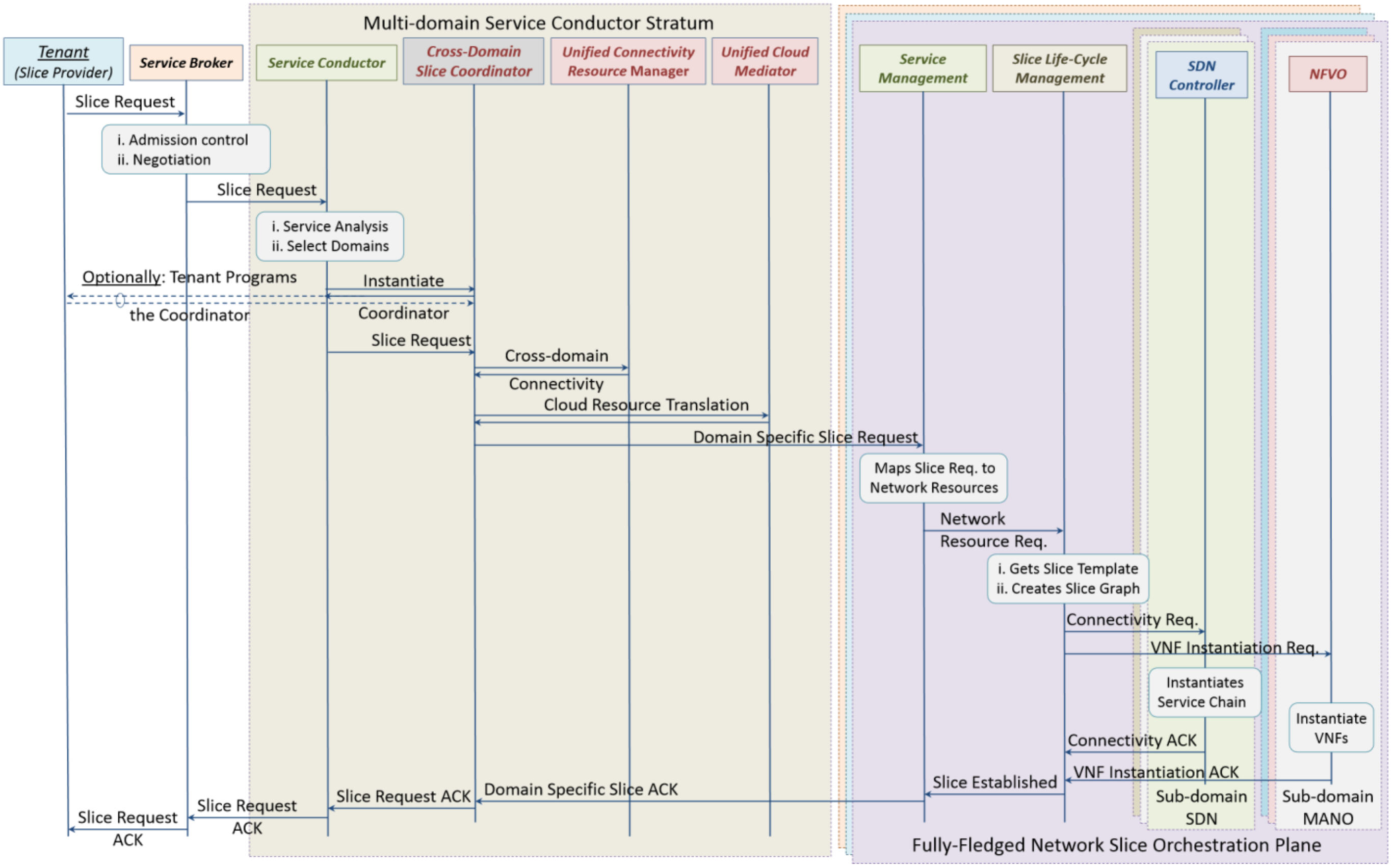

Figure 4: Sequence diagram for creating a multi-domain slice.

The Slice Life-Cycle Management Function selects the appropriate slice template and creates the desired "slice resource graph". It then carries out the resource configuration towards the corresponding sub-domain by issuing a request towards the respective Sub-domain NFV MANO and/or Sub-domain-specific SDN Controller, which in turn needs to create the desired NFV, computing and connectivity slate. There are two major options when configuring an NFV or computing slate: (i) the Sub-domain NFVO forwards the request directly to the corresponding VIM or (ii) it communicates the request to related service chain. A multi-domain NSI becomes operational when all domain-specific NSSIs and cross-domain connectivity are configured successfully. Once the resources are granted, an acknowledgement shall be returned to the tenant, updating also the Service Broker.

*1) Scenario I: Multi-domain Resource Modification:* A resource modification request typically concerns a particular slate and is handled within the corresponding Sub-domain NFVO or SDN controller via conventional mechanisms that scale up or down VNF resources or perform routing alternations. When a modification request relies on resource re-configuration beyond the capabilities of a sub-domain, e.g. a certain VNF cannot be scaled up further from the same sub-domain but instead can be configured into a different one; the connectivity and VNF reconfiguration would be handled by the Slice Life-Cycle Management Function.

Figure 5a provides an overview of such resource modification procedure. A new resource allocation can then be determined with the Slice Life-Cycle Management Function instructing the corresponding sub-domain SDN controller and/or NFVO about the related modifications that need to take place. The NFVO instantiates/terminates or modifies the indicated VNF(s). It instructs the VNFM and VIM to carry out the corresponding re-configurations or resource scaling up/down including the reclamation of unused resources. The SDN controller then updates the service chain, providing an acknowledgement all the way back.

*2) Scenario II: Multi-domain Service Modification:* In case of an unsuccessful resolution, the Slice LifeCycle Management Function invokes the Service Management to check other potential resource mapping for re-assigning VNFs and connectivity on different subdomains. Such a process can lead the Slice Life-Cycle Management Function to re-assign a "slice resource graph" that may result in a different resource allocation across sub-domains. If the Service Management function fails to identify a valid resource re-mapping, then a single domain alone cannot handle the slice modification request and hence the Cross-domain Slice Coordinator needs to re-assign the allocated resources differently across the federated domains.

The Cross-domain Slice Coordinator is a federated NSI manager, which assigns logical resources, while performing resource monitoring and control considering the desired performance targets. Once a modification request cannot be handled by rearranging the allocated resources among the involved sub-domains, the Crossdomain Slice Coordinator instructs the Multi-domain Service Conductor to modify the service realization involving optionally other domains not previously utilized with the guidance of the Service Broker. Figure 5b illustrates the main processes for modifying the allocated services as a response to a modification request.

The Multi-domain Service Conductor analyzes once again the service requirements with the objective to decompose a slice request across a different set of federated domains. Once this is accomplished, it informs the Cross-domain Slice Coordinator about the modified service mapping. The Cross-domain Slice Coordinator identifies the type of modification with respect to

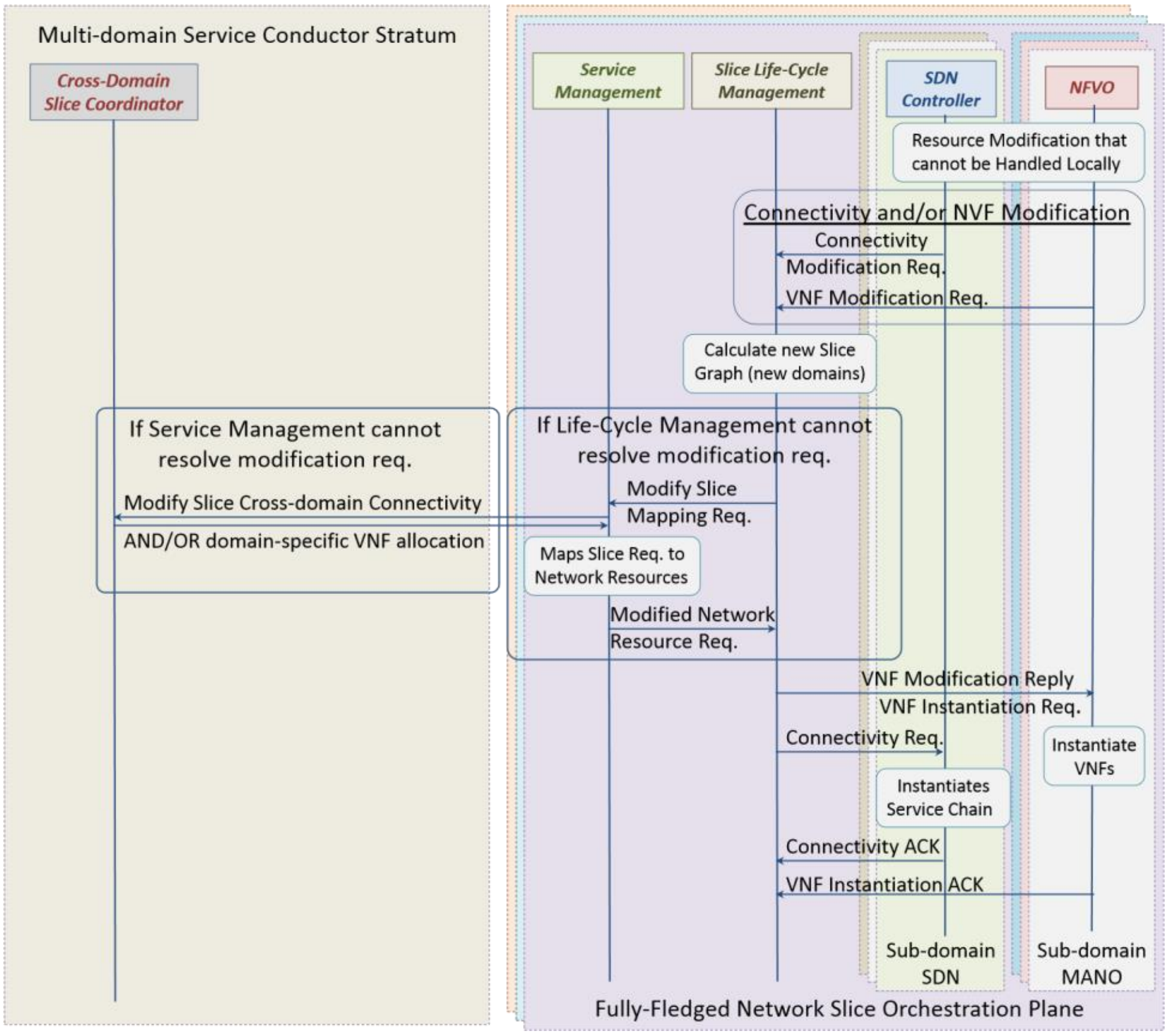

(a) Scenario I: Sequence diagram for updating the federated resources of a multi-domain slice.

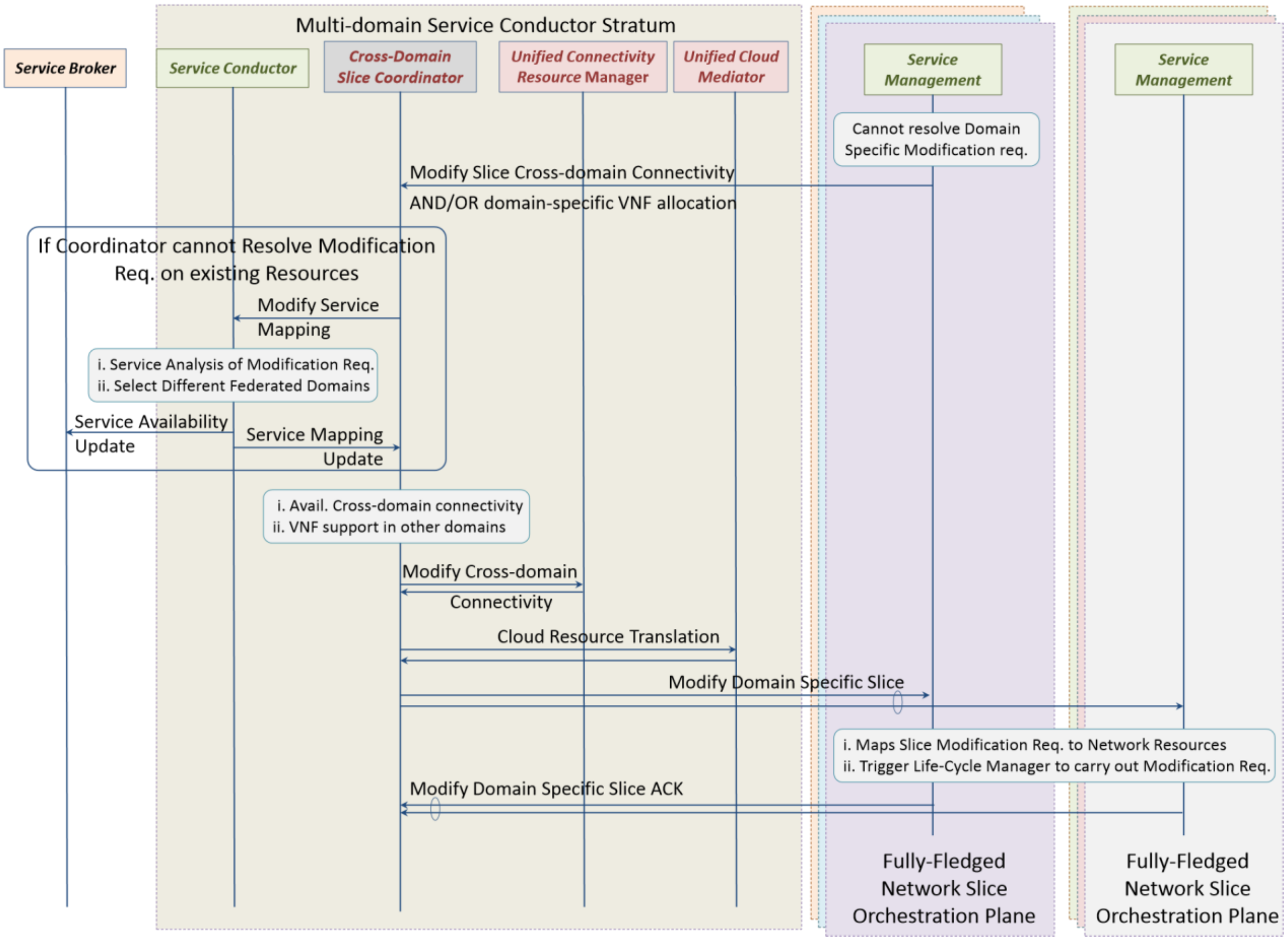

(b) Scenario II: Sequence diagram for updating the service across multiple domains.

Figure 5: Multi-domain slice modification procedures

particular domains, i.e. allocate new, scale up/down, or terminate resources, relying on the Unified Cloud Mediator and Unified Connectivity Resource Manager. The Cross-domain Slice Coordinator then provides a domain specific slice modification request to the Service Management function. Each Service Management function in turn performs a mapping analysis to identify the desired modifications or allocation of network resources that correspond to certain sub-domains. It then triggers the Life Cycle Manager to carry out the related modification requests by involving the respective Subdomain NFV MANO and/or SDN Controller. Once the desired modifications take place, an acknowledgement is returned to each Service Management function and then to the Cross-domain Slice Coordinator.

## V. Discussion & Open Challenges

Multi-domain NS orchestration and management has not been fully explored with various deployment specific issues being still open. Herein, we explore three fundamental emerging aspects relevant to service management interfaces, resource isolation and sharing, and service based management plane.

### *A. Service Management Interfaces & Service Profiling*

The wide-adoption of slicing relies on standardized interfaces and relevant information models, which can abstract service capabilities and resource requirements hiding the network specifics. Currently, RESTfull models can be used by 3*rd* parties, e.g., verticals, for programmability purposes, facilitating also information exchange and control among different technology and/or administrative domains. A number of information models exists and are currently under development to convey transport and cloud capabilities towards the mobile network management plane including

L3SM[3]/L2SM[4] and NFVIFA Os-Ma-Nfvo[5]. Although such data models aim to enrich the NS management plane, they are not taking into account: (i) multi-domain connectivity and control considerations, neither (ii) resiliency and performance measurements on federated resources.

Besides the development of data models, when acquiring cloud and/or networking resources, it is essential to analyze and map the service requirements of the corresponding slice into relevant resources. A NSI may/may not serve one type of traffic, while the connectivity and resource demand may not be equally distributed among all indicated end-points. Service profiling algorithms are hence needed to optimize the mapping of allocated resources especially for federated environments.

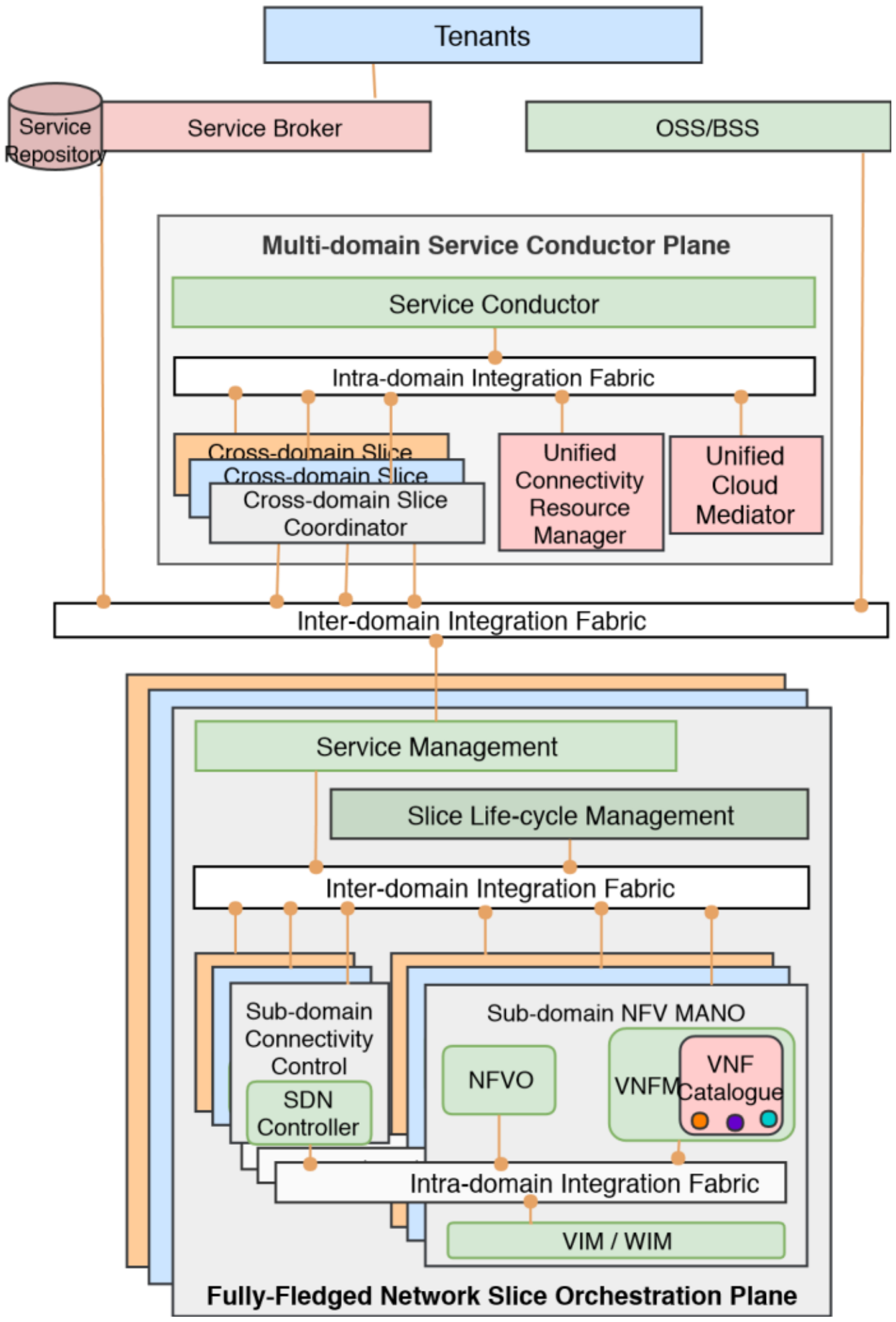

Figure 6: A service based management architecture for multi-domain slicing.

## *B. Resource Sharing & Isolation*

The notion of resource sharing and isolation has a further significance when considering federated domains. Selecting which functions and connectivity resources shall be shared or kept dedicated impacts the end-toend performance and the economic cost. Realizing a common control plane and linking it to a dedicated one per slice considering cross-domain resources taking into account, e.g. latency and resource utilization, is yet to be explored. Security is another isolation-relevant issue for configuring and operating network slices with federated resources. Two security aspects should be considered including authorization and encryption. Cross-domain security can be carried out by extending border security protocols among different administrative domains in coordination with the multi-domain service management.

## *C. A Service Based Network Management Architecture*

A service based architecture [16] relies on a communication bus that offers function inter-connectivity instead of point-to-point interfaces. A mediator, referred to as function repository, assists registered functions to cooperate based on service needs by allocating a light-weighted interface via the communication bus. In this way management services can be modified independently, i.e. being modular, with minimal impact among each other and management capabilities can be customized considering the needs of a particular NSI.

A preliminary service based architecture vision for multi-domain slicing is illustrated in Figure 6 following the Zero touch network and Service Management (ZSM) paradigm[6]. An intra-domain bus is envisioned to connect orchestration and control functions per slate. Different technology slates including the related service and resource management can then interact via an inter-domain integration fabric forming a Fully-Fledged NS orchestration plane. In each multi-domain NSI, a cross-domain slice orchestrator manages a NS service combining diverse Fully-Fledged orchestration planes related with the RAN, transport and core network via an inter-domain fabric. In addition, it interacts via an intradomain bus internally within the multi-domain service conductor plane for carrying out service management and configuration procedures. Certain capabilities and operational details of such architecture are still open, including the function repository, the organization of data management states and the notion of stateless management, where the processing and storage is separated. The interaction with the control plane and the integration of data analytics are yet further issues to be explored.

[3] Q. Wu, S. Litkowski, L. Tomotaki, K. Ogaki, YANG Data Model for L3VPN Service Delivery, IETF RFC 8299, Jan. 2018.

[4] B. Wen, G. Fioccola, C. Xie, L. Jalil, A YANG Data Model for Layer 2 Virtual Private Network (L2VPN) Service Delivery, IETF RFC 8466, Oct. 2018

[5] ETSI GS NFV-IFA, Os-Ma-Nfvo reference point, Interface and Information Model Specification, Oct. 2016

[6] https://www.etsi.org/technologies-clusters/technologies/zerotouch-network-service-management

## VI. Conclusions

This paper elaborates a multi-domain orchestration and management architecture and framework to address the service challenges of network slicing when utilizing federated resources. In particular, a multi-domain Service Conductor plane is introduced considering: (i) its main functional components including the Cross-domain Slice Coordinator and its assisting Unified Connectivity Resource Manager and Unified Cloud Mediator elements, and (ii) inter-working issues with the conventional single administrator Fully-Fledged network domain, wherein NSSIs are established by combining computing, storage and network slates with RAN, transport and core network capabilities. The main operations are elaborated considering a multi-domain NSI instantiation and management, bringing also an insight into the further architectural and operational challenges.

## Acknowledgment

This work was partially supported by the European Union's Horizon 2020 Research and Innovation Program through the 5G!Pagoda project and the MATILDA Project with Grant No. 723172 and No. 761898 respectively. It was also supported in part by the 6Genesis project under Grant No. 318927.